\documentclass[aps,prb,superscriptaddress,twocolumn]{revtex4}
\usepackage{bm}
\usepackage[fleqn]{amsmath}
\usepackage{hyperref}
\usepackage{graphicx}

\begin{document}

\title{Surface States of a System of Dirac Fermions: A Minimal Model}

\author{V.A. Volkov}
\email{Volkov.v.a@gmail.com}
\author{V.V. Enaldiev}
\affiliation{Kotel'nikov Institute of Radio-engineering and Electronics of the Russian Academy of Sciences, 11-7 Mokhovaya St, Moscow, 125009 Russia}

\date{\today}

\begin{abstract}
A brief survey is given of theoretical works on surface states (SSs) in Dirac materials. Within the formalism of envelope wave functions and boundary conditions for these functions, a minimal model is formulated that analytically describes surface and edge states of various (topological and nontopological) types in several systems with Dirac fermions (DFs). The applicability conditions of this model are discussed.
\end{abstract}

\maketitle

\section{THE ENVELOPE-FUNCTION METHOD. INTRODUCTION TO THE HISTORY OF THE PROBLEM}
By the middle of the 20-th century, the problem arose of explaining and predicting the electronic properties of semiconductors in external fields. Various versions of the single-band effective-mass method were developed. The Luttinger--Kohn envelope-function (EF) method \onlinecite{Luttinger_1955} based on a generalization of the $\bm{kp}$-approach turned out to be very convenient. The formalism of EFs admitted a natural multiband generalization. Such a generalization was made by Keldysh in his theory of deep impurities \onlinecite{Keldysh_1964}. In [\onlinecite{Keldysh_1964}], Keldysh also noticed that, under certain conditions, EFs in a narrow-band semiconductor obey an effective Dirac equation (a system of four first-order differential equations). It is interesting that, in III--V semiconductors, this situation occurs under reversal of the sign of the strong spin--orbit splitting of the valence band. The idea of inversion of bands in crystals with strong spin--orbit interaction will be addressed below in this paper. In the same year as [\onlinecite{Keldysh_1964}], it was shown [\onlinecite{Wolff_1964}] that the spectrum of electrons and holes in bismuth near the $L$ points of the Brillouin zone should be described by an anisotropic Dirac Hamiltonian. 

According to modern terminology, bismuth can be considered as a Dirac material, and the first Dirac material at that. These materials include graphene, bismuth-- antimony alloys, lead chalcogenides, 2D and 3D topological insulators, Dirac semimetals, Weyl semimetals, and many other materials. One-particle excitations in these materials are called massless (for gapless materials) or massive Dirac fermions (DFs). 

The energy spectrum $E(p)$ of free DFs in the three-dimensional isotropic case is analogous to the spectrum of a relativistic electron: 
\begin{equation}\label{Dirac_spectrum}
E(p)=\pm\sqrt{m^2c^4 + p^2c^2}
\end{equation}
Here $m$ is the electron effective mass, which is usually one or two orders of magnitude less than the mass of a free electron in vacuum, and $c$ is the effective velocity of light. This velocity is two orders of magnitude less than the velocity of light in vacuum; therefore, real Fermi excitations are, in fact, non-relativistic. However, relativistic corrections due to spin-orbit interaction in crystals with close bands may be rather large, which contributes to the formation of DFs. A finite mass in the spectrum of relativistic particles corresponds to a finite width of the forbidden band according to the well-known formula $E_g = 2mc^2$. 

The 1960s marked the emergence of the physics of 2D electron systems. Quantum size effect was started to be used to realize the 3D $\to$ 2D transition. The conventional theory of size quantization is based on the single-band effective mass approximation with zero boundary conditions for EFs. Physically, this is justified by the impenetrability of potential barriers on the surfaces (interfaces) that confine the motion of an electron. This approximation works well in crystals with parabolic band spectrum, for example, for electrons in silicon. However, the situation is qualitatively changed when one tries to correctly describe the size quantization in a system of DFs. 

The possibility of implementation of the size quantization was predicted in 1962 [\onlinecite{Sandom_1962}] for films made of narrow-band semiconductors and related semimetals with low concentration of carriers. The experimental discovery of this phenomenon [\onlinecite{Ogrin_1966}, \onlinecite{Lutskii_1966}] and its explanation [\onlinecite{Sandom_1967}] occurred a few years later when studying bismuth films. 

Strictly speaking, envelope wave functions strongly differ from real wave functions of an electron in a crystal: the former functions are the envelopes of the latter. However, this fact was not of fundamental importance for qualitative considerations at the initial stage. But the development of experiment raised the question of quantitative description of the results. Unexpectedly, the problem turned out to be rather difficult. A seemingly academic question of the simplest boundary conditions for effective wave functions on the surface of a bismuth-type narrow-band semiconductor inevitably entailed the problem of description of SSs. It turned out that the simplest and physically obvious boundary conditions---zero boundary conditions for all four EFs in the Dirac equation---lead to a trivial (everywhere zero) solution due to the overdeterminacy of the problem. This is how the theoretical problem of derivation of correct boundary conditions for the envelope-function method arose, which is especially important in its multiband version. It is important that these boundary conditions should also describe the spectrum of SSs in addition to volume states. 

The point is that, on the real surface of any crystal, including a Dirac crystal, there always exist electron or hole SSs, which are extrinsic and/or intrinsic. The first are attributed to defects and contamination of the surface, while the second (we will consider only these SSs) exist on the ideal surface. Intrinsic SSs were theoretically considered at the early stage of band theory in different models [\onlinecite{Tamm_1932_1,Tamm_1932_2,Maue_1935,Shockley_1939,Davison_1970}]. I.E. Tamm's works were the first ones; therefore, these SSs are called Tamm states. W. Shockley considered another popular model, and the SSs obtained by him are often called Shockley states. However, there is no essential difference between the Tamm and Shockley SSs. Moreover, in the simplest models, these states turn into each other under the variation of model parameters [\onlinecite{Davison_1970}], which may lead to confusion. Therefore, we will call intrinsic SSs Tamm--Shockley states, bearing in mind that both types of SSs are attributed to the sharp (at the atomic scale) discontinuity of the crystal potential at the surface. 

The Tamm--Shockley states are not only situated (completely or partly) in the forbidden band of a bulk crystal. It is very important that these states should form a surface band of conducting states that are delocalized in the plane of the surface and are characterized by a 2D dispersion law. However, intrinsic SSs are very sensitive to surface roughness and contamination. Therefore, it is not surprising that a lot of time passed until the Tamm--Shockley states were reliably detected experimentally. As a rule, these states are investigated in ultra-high vacuum usually by local techniques (STM, ARPES, and so on). The problem of existence of an SS band under normal conditions requires special investigation. For instance, it is not quite clear how much stringent conditions should be imposed on the perfection of the surface or the interface, and what does this depend on. 

In ordinary semiconductors (with wide forbidden gap and small spin--orbit interaction), a band of Tamm--Shockley states arises far from always, and the wave functions of these states are not usually mixed with the wave functions of size quantized states. The very existence conditions of these states strongly depend on the structure of the surface or the interface at the atomic scale; therefore, in spite of the 80-year history of the problem, the general principles of behaviour of these functions have been investigated rather fragmentarily, especially experimentally. The situation started to be changed in recent years, when topological materials described by a modified Dirac equation appeared. The theory of topological insulators predicts that "topological" SSs protected from backscattering should exist in these materials for topological reasons. The very fact of existence of these states does not depend on the details of the structure of a surface region.

A more general question arises: are the requirements imposed on the perfection of the surface (interface) relaxed for the manifestation of SSs on the surface of any Dirac crystal, not necessarily a topological insulator? The theoretical answer in the simplest Dirac model [\onlinecite{Volkov_Pinsker_1981_English}] is positive. Nevertheless, as expected for the Tamm--Shockley states, the spectrum of SSs for Dirac materials depends on the properties of the surface, although not in a quite obvious way. 

To solve this problem, one had to derive boundary conditions of general form that are invariant under time reversal and describe DFs near an impenetrable wall. The solution of the genuine (rather than a modified, as in the theory of topological insulators later) Dirac equation in a half-space with these boundary conditions showed that SSs should appear on any surface (more precisely, for any parameters of the boundary conditions) and should be extremely strongly split with respect to spin as a result of spin--orbit interaction with the surface (this interaction is often called the Rashba interaction). These SSs are characterized by a conical dispersion law but they occurred not all the values of momenta. The "strength" of surface spin--orbit interaction is characterized by a real phenomenological parameter $a_0$ that appears in the boundary conditions. Depending on the sign of $a_0$, there exist two classes of surfaces. For the surfaces of one of the classes, the conical ("Dirac") point in the spectrum is located in the forbidden gap of the bulk material; i.e., SSs in this case have a 2D spectrum of massless spin-nondegenerate DFs, as in the case of a topological insulator. The physical meaning of the sign of $a_0$ remained unclear. 

In 1985, in the famous work [\onlinecite{Volkov_Pank_1985_English}], the authors considered another Dirac model: the model of inverse heterocontact. It is described by the Dirac equation in which the mass of a DF is varied in space smoothly (at the atomic scale) and symmetrically with respect to the electron--hole (e--h) transformation. The interface is determined by the position of the plane in which the sign of the mass is reversed (the conduction and valence bands are interchanged). It is this case where the band of nondegenerate heterointerface states possessing a conical 2D spectrum of massless DFs is formed. It is remarkable that this conclusion does not depend on the details of the interface potential; one just needs the inversion of the bands. The point is that, under these conditions, the equation for DFs in the inverse contact takes a form typical for supersymmetric quantum mechanics with zero mode corresponding to interface states. This nontrivial result is actively used in modern physics of topological insulators. The Dirac point in this symmetric model is located exactly at the center of the gap. 

The reason for some disagreement between the results of [\onlinecite{Volkov_Pank_1985_English}] and [\onlinecite{Volkov_Pinsker_1981_English}] and the physical meaning of the sign of $a_0$ were elucidated in the survey [\onlinecite{Volkov_review_1995_English}]. A significant e--h asymmetry was introduced into the model of an abrupt heterocontact (which, however, is still smooth at the atomic scale). As this asymmetry increases, the Dirac point was more and more shifted from the center of the gap and disappeared. For a greater discontinuity of bands at the heterojunction, the spectrum of interface states completely corresponds to the spectrum of SSs from [\onlinecite{Volkov_Pinsker_1981_English}]. The comparison of the results yielded a model expression for the boundary parameter $a_0$. The sign of this parameter generally correlated to the reversal of the sign of the gap at the interface. 

Unfortunately, the incipient state of the manufacturing technology of samples did not allow one to verify these conclusions experimentally 30--20 year ago. The works [\onlinecite{Volkov_Pinsker_1981_English,Volkov_Pank_1985_English,Volkov_review_1995_English}] acquired importance in relation to the discovery of topological insulators, as well as in relation to the recently detected conducting edge states of non-topological type in nanoperfortated graphene [\onlinecite{Latyshev_2013}, \onlinecite{Latyshev_SciRep_2014}]. 

In the present study, we discuss the problems touched upon above and, within the formalism of EFs, formulate a minimal model that describes SSs of various, topological and non-topological, types in a number of Dirac materials. The paper is organized as follows. In Section 2, we give a brief survey of works on Tamm--Shockley-type SSs within the formalism of EFs. In Section 3, we consider topological SSs and, in Section 4, Tamm--Shockley edge states in graphene. In Section 5, we formulate conclusions.

\section{TAMM--SHOCKLEY-TYPE SURFACE STATES}
The method of effective wave-functions---envelope functions---is widely used to describe the behavior of electrons in multilayer semiconductor structures. The method of EFs can be applied to the description of smooth (at the atomic scale) fields and is inapplicable to the real case of atomically abrupt interfaces. Information on the microscopic structure of the interface can be taken into account by appropriate boundary conditions for EFs. 

The problem of boundary conditions in bounded crystals has a long history. Theoretical studies on boundary conditions in semiconductor structures can be classified into two groups. The works of the first, largest, group are devoted to the derivation of "two-sided" boundary conditions relating EFs to their left and right derivatives at the interface. They contain different approaches to the solution of mathematical problems related, in particular, to possible singular behaviour of EFs on a heterointerface [\onlinecite{Ivchenko2005,Zawadzki2004,Foreman_2005,Takhtamirov_1999,Rodina_2002,
Takhtamirov_1997,Takhtamirov_2010}]. 

We will focus only on the works of the second group, which are devoted to the derivation of one-sided boundary conditions at the crystal--high barrier interface (in particular, at the crystal--vacuum interface). The problems of this type arise when describing the Tamm--Shockley surface (interface) states. 

Under the neglect of the spin--orbit interaction, a microscopic derivation of boundary conditions for EFs at a step-like boundary between semiconductor ($z > 0$) and vacuum ($z < 0$) was apparently first presented in [\onlinecite{Volkov1976}, \onlinecite{Volkov1977}]. The boundary conditions introduced there contain boundary parameters that are analytically (but in a complicated way) expressed in terms of a complete infinite-band structure of a semiconductor analytically continued to the domain of complex quasimomenta. The numerical determination of these parameters remains an unsolved problem. 

\section{SHALLOW TAMM--SHOCKLEY STATES IN A SINGLE-BAND APPROXIMATION}

In a single-band approximation, a boundary condition represents a linear relation between an EF and its normal derivative with a single boundary parameter with the dimensionality of length, which we denote below by $R$. This length characterizes the structure of the interface between semiconductor ($z > 0$) and an impenetrable barrier ($z < 0$) at the atomic scale and has the meaning of the localization depth of a shallow Tamm--Shockley state when it exists (to this end, the condition $R > 0$ should be satisfied). Moreover, the length $R$ depends on the parameters of the bulk band structure. 

In [\onlinecite{Volkov_1979}], a much simpler derivation of the same boundary condition is presented from the Hermiticity of the effective Hamiltonian for EFs in a half-space bounded by an impenetrable barrier. Within such a phenomenological approach, the parameter $R$ should be determined from experiment. The high-barrier model is applicable when the interface length $R$ is large compared with the penetration depth into the barrier. The effect of the spin--orbit interaction on single-band boundary conditions and spin splitting of shallow Tamm--Shockley states in the conduction band of a semiconductor with inversion symmetry was considered in [\onlinecite{Vasko1979}]. In the model used in [\onlinecite{Vasko1979}], the spin splitting is controlled by the product of the length $R$ and the parameter of bulk spin--orbit interaction. A non-parabolic generalization of boundary conditions [\onlinecite{Vasko1979}] in an asymmetric quantum well with infinite barriers is presented in [\onlinecite{Rodina_2006}]. 

We begin with the analysis of a single-band boundary condition as applied to heterostructures based on III--V semiconductors with the heterointerface orientation (001). In [\onlinecite{Devizorova_2013},\onlinecite{Devizorova_2014}], the authors considered the effect of an atomically abrupt heterointerface on the effective single-band Hamiltonian of a quantum well and the spin splitting in a conduction band of symmetry $\Gamma_{6c}$ in crystals with lack of inversion symmetry. The discontinuity of bands at the heterointerface is assumed to be large, and the heterobarrier, to be impenetrable. The latter is characterized by a certain boundary condition for EFs. 

The dynamics of a conduction electron for $z > 0$ within the multiband EF method is described by the system of Kohn--Luttinger $\bm{kp}$-equations 
\begin{eqnarray}\label{kp_eqs}
\left\{ \left[E_n(0)+\frac{\hat{\bm{p}}^2}{2m_0}+V(z)\right]\delta_{nn'} + \frac{\hat{\bm{p}}\cdot\bm{p}_{nn'}}{m_0} \right. \nonumber \\
+ \left. \frac{\hbar}{4m_0^2c^2}\left(\bm{p}\cdot\left[\bm{\sigma}\times\bm{\nabla}V_0\right]\right)_{nn'} \right\}\Phi_{n'}=E\Phi_n
\end{eqnarray}
where $n$ is the band number, $E_{n}(0)$ is the extremum energy of the $n$-th band, $\Phi_n$ is a set of EFs, $\bm{p}_{nn'}$ is a matrix element of the momentum operator $\hat{\bm{p}}$ on Bloch functions at the center of the Brillouin zone, $m_0$ is the mass of a free electron, the last term in curly brackets is a matrix element of the operator of bulk spin--orbit interaction on Bloch functions, and $\sigma_i$, $i = x, y, z,$ are Pauli matrices. 

The Hermiticity of Hamiltonian (\ref{kp_eqs}) in the half-space $z > 0$ reduces, after integration by parts, to vanishing of the surface contribution. This is equivalent to setting the matrix element of the normal component of the current operator between any pair of states at the boundary to zero: 
\begin{equation}\label{normal_current}
\left.\left(\Phi^{+}_{\lambda}\hat{v}_z\Phi_{\nu}\right) \right|_{z=0}=0
\end{equation}
where $\hat{v}_z$ is a non-diagonal matrix of velocity $\left (\hat{v}_z\right)_{nn'}= \partial_{p_z} \left(H_{nn'}\right)$. 

Since only the spinor corresponding to the conduction band $\Gamma_{6c}$ is large in the multicomponent function $\Phi$, we apply the unitary transformation $\Phi = e^S\phi$ [\onlinecite{Winkler}] (with regard to $\bm{kp}$ terms up to the third order inclusively), which reduces the Hamiltonian to a single-band one with effective mass $m*$ and smooth potential $V(z)$. Now, the 3D Hamiltonian of the conduction band contains contributions $\hat{H}_{BIA}$ and $\hat{H}_{SIA}$ that describe the spin splitting due to the lack of inversion symmetry of the crystal and the asymmetry of the well: 
\begin{equation}\label{hamiltonians}
\hat{H} = \frac{\hat{\bm{p}}^2}{2m^*}+V(z)+\hat{H}_{BIA}+\hat{H}_{SIA},
\end{equation}
\begin{eqnarray}
\hat{H}_{BIA} =\frac{\gamma_c}{\hbar^3}\left[ \sigma_xp_x\left(\hat{p}_y^2 - \hat{p}_z^2\right)\right. \qquad\qquad\qquad\quad\nonumber \\
+\sigma_yp_y\left(\hat{p}_z^2 - \hat{p}_x^2\right) + \sigma_zp_z\left(\hat{p}_x^2 - \hat{p}_y^2\right),
\end{eqnarray}
\begin{equation}
\hat{H}_{SIA} = a_{SO} \left(\sigma_xp_y - \sigma_y p_x\right)\partial_zV(z).
\end{equation}
By the same transformation, condition (3) reduces to a certain constraint for a two-component EF. Next, we require the invariance of this constraint with respect to the operation of time reversal 
\begin{equation}\label{time_inversion}
\hat{T} = i\sigma_y \hat{K},
\end{equation} 
where $\hat{K}$ is the operator of complex conjugation. We obtain T-invariant boundary conditions that take into account the spin--orbit interaction in the bulk and on the symmetry interface $C_{2v}$, as well as the absence of the inversion center in the bulk crystal: 
\begin{eqnarray}\label{single_band_BC}
\left[\sigma_0 - i\frac{R\hat{p}_z}{\hbar} - i \frac{2m^*\gamma_c R}{\hbar^4}\left(\sigma_yp_y - \sigma_xp_x\right)\hat{p}_z\right. \nonumber \\
- i \frac{m^*\gamma_c R}{\hbar^4}\sigma_z\left( p_x^2 - p_y^2\right) + \frac{\left(\chi + \chi^{int}\right)R}{\hbar}\left(\sigma_xp_y - \sigma_y p_x\right) \nonumber\\
\left.\left.- \frac{2m^*\gamma_c^{int}}{\hbar^3}\left(\sigma_yp_y - \sigma_x p_x\right) \right]\phi\right|_{z=0}=0.\qquad
\end{eqnarray}
Here $\sigma_0$ is the identity matrix. The real quantity $R$, which has the dimensionality of length, depends on the microscopic structure of the boundary. The constants $\gamma_c$ and $\chi$ are defined by the volume parameters; for GaAs, $|\gamma_c| = 24.4$ eV$\cdot {\rm A}^3$ and $\chi  = 0.082$. The constants $\gamma_c^{int}$ and $\chi_c^{int}$ characterize the spin--orbit interaction with the interface crystal potential and are determined from a comparison with experiment [\onlinecite{Devizorova_2014}]. 

The Tamm problem in the half-space $z > 0$ in this case corresponds to the solution of a single-band effective mass equation with Hamiltonian (\ref{hamiltonians}) and boundary conditions (\ref{single_band_BC}). It is important that such an approach leads to the discontinuity of single-band EFs on the interface. This is attributed to the non-perturbative effect of the interface potential. Therefore, the approach allows one to describe, for example, shallow Tamm--Shockley states, if they exist, even in the single-band approximation [\onlinecite{Volkov1976,Volkov1977,Volkov_1979}]. Unfortunately, this situation is not realized for electrons of the conduction band at the (001) GaAs/AlGaAs hererojunction because $R < 0$. A quantitative comparison with precision measurements of electron spin resonance that are sensitive to the parameters of boundary conditions shows [\onlinecite{Devizorova_2014}] that $R = -2.2$ nm. 

%%%%%%%%%%%%%%%%%%%%%%%%%%%%%%%%%%%%%%%%%%%%%%%%%%%%%%%%%%%%%%%
\begin{figure}
\includegraphics{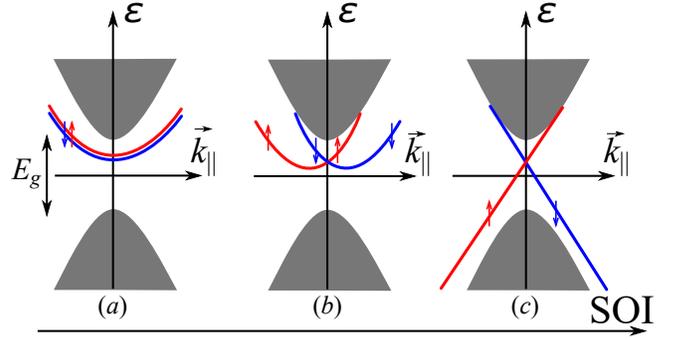}
\caption{Evolution of the 2D spectrum of Tamm--Shockley SSs for increasing interface spin--orbit interaction: from (a) spin-degenerate shallow SSs through (b) weak spin--orbit splitting to (c) a conical spectrum of topological-type SSs. Bulk states are shown in gray. }
\end{figure}
%%%%%%%%%%%%%%%%%%%%%%%%%%%%%%%%%%%%%%%%%%%%%%%%%%%%%%%%%%%%%%%%%

On the surface or at the interface where $R > 0$, a band of shallow SSs [\onlinecite{Volkov_1979}] is formed that is described, with neglect of the spin--orbit interaction, by the first two terms in the boundary conditions (\ref{single_band_BC}) (Fig. 1a). These SSs are "suspended" over the bottom of the conduction band, and their localization depth is equal to $R$. The volume and interface contributions to the spin--orbit interaction, which are described by the remaining terms in (\ref{single_band_BC}), anisotropically split the band of SSs with respect to spin (more precisely, with respect to chirality---the projection of spin onto the direction related to momentum) (Fig. 1b). This splitting was first introduced in [\onlinecite{Vasko1979}] in the simplest isotropic model. As the interface spin--orbit interaction increases, the splitting increases and becomes comparable to the bandgap $E_g$ (Fig. 1c), and the single-band approximation fails. 

\section{TAMM--SHOCKLEY STATES IN THE TWO-BAND APPROXIMATION}

The two-band model is a minimal multi-band model. For Dirac materials such as bismuth, lead chalсogenides, and many other, the Tamm problem in the simplest approximations reduces to the solution of the Dirac equation in the half-space $z > 0$, 
\begin{equation}\label{Dirac_eq}
\left(
\begin{array}{cc}
mc^2 - E & c \bm{\sigma}\cdot\bm{k} \\
c \bm{\sigma}\cdot\bm{k} & -mc^2 - E
\end{array}
\right)
\left(
\begin{array}{cc}
\Psi_c \\
\Psi_v
\end{array}
\right)=0
\end{equation}
with the boundary conditions [\onlinecite{Volkov_Pinsker_1981_English}]
\begin{equation}\label{Dirac_BC}
\left[\Psi_v - ia_0\bm{\sigma}\cdot \bm{n}\Psi_c\right]_S=0.
\end{equation}
Here the two-component EFs $\Psi_c$ and $\Psi_v$ describe the conduction and valence bands, $a_0$ is a real phenomenological parameter describing the small-scale structure of the surface, $\bm{n}$ is a unit normal to the surface $S$ of the crystal, and $E_g = 2mc^2$ is the bulk band gap. The boundary conditions (\ref{Dirac_BC}) follow from the Hermiticity (self-adjointness) of the Dirac Hamiltonian in the half-space and the symmetry of the problem with respect to time reversal. 

%%%%%%%%%%%%%%%%%%%%%%%%%%%%%%%%%%%%%%%%%%%%%%%%%%%%%%%%%%%%%%%
\begin{figure}
\includegraphics{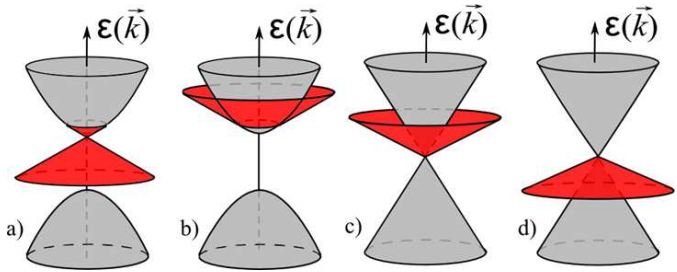}
\caption{Energy spectrum of the 3D Dirac equation in a half-space. The band of non-degenerate Tamm--Shockley states is shown in red, and bulk states are shown in light gray. There are two classes of surfaces numbered by the sign of the boundary parameter $a_0$. (a) An analog of a topologically nontrivial spectrum. For a surface with positive $a_0$, SSs fill the main part of the cone with a Dirac point in the gap of a bulk material, as in a topological insulator. (b) An analog of a topologically trivial spectrum. As the sign of $a_0$ is changed, SSs are expelled from the bulk gap and fill the peripheral part of the cone. (c, d) The spectrum of a single-valley Dirac semimetal is obtained from the spectrum of the Dirac equation for zero mass. The massless limit of the Tamm problem for two classes of surfaces is shown in Figs. 2d and 2c; Figures 2a and 2b turn into Figs. 2d and 2c, respectively.  }
\end{figure}
%%%%%%%%%%%%%%%%%%%%%%%%%%%%%%%%%%%%%%%%%%%%%%%%%%%%%%%%%%%%%%%%%

A perturbative account of remote bands gives rise to diagonal corrections to the Dirac Hamiltonian in (\ref{Dirac_BC}) that are quadratic in momentum. The corresponding band gap model, sometimes called Dimmock's model, is used in the theory of topological insulators (see below). In this section, the contribution of remote bands is neglected. 

As a result of the strong spin--orbit interaction, the Tamm problem for DFs always gives rise to a non-degenerate spectrum of SSs that fills a part of the cone surface (Fig. 2). It is interesting to compare this result with the theory of topological insulators and semi-metals. The result qualitatively depends on the sign of the boundary parameter $a_0$. 

For a surface with a positive value of $a_0$, a conical (''Dirac'') point in the spectrum lies in the forbidden gap of a bulk single-valley material; i.e., in this case, SSs are characterized by a 2D spectrum of massless DFs (Figs. 1c and 2a). The comparison with topological theory implies that this case is an analog of a topologically nontrivial case. Figure 2b corresponds to the topologically trivial phase, for which the parameter $a_0$ is negative. In the massless limit, we obtain SSs for Dirac's single-valley 3D semimetal (Figs. 2c, 2d). 

%%%%%%%%%%%%%%%%%%%%%%%%%%%%%%%%%%%%%%%%%%%%%%%%%%%%%%%%%%%%%%%
\begin{figure}[h]
\includegraphics{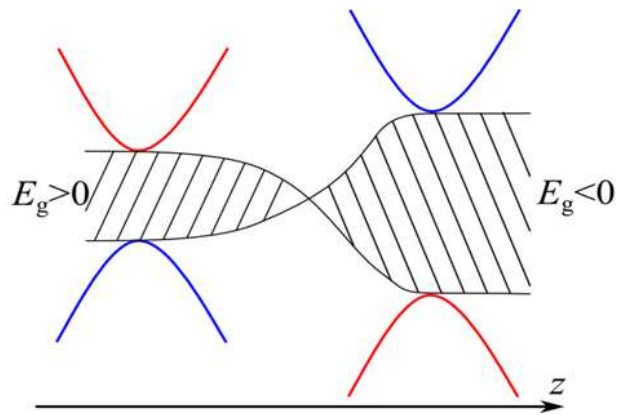}
\caption{Dirac model of a symmetric heterocontact [\onlinecite{Volkov_Pank_1985_English}] in the case of reversal of the sign of interband gap is described by equations of supersymmetric quantum mechanics; the zero mode of these equations corresponds to interface states with conical spectrum (Fig. 1c) that are stable with respect to perturbations. }
\end{figure}
%%%%%%%%%%%%%%%%%%%%%%%%%%%%%%%%%%%%%%%%%%%%%%%%%%%%%%%%%%%%%%%%%

Analytically, the spectrum of SSs for the Dirac equation in the half-space is expressed as 
\begin{equation}\label{SS_spectrum}
E_s(\bm{k}_{||})=s\frac{2a_0}{1+a_0^2}|\bm{k}_{||}|+E_0.
\end{equation}
Here the quantum number $s = \pm 1$ describes the chirality, $\bm{k}_{||} = (k_x, k_y, 0)$. The energy of the conical point is measured from the center of the gap and has the form 
\begin{equation}\label{Dirac_point}
E_0=mc^2\frac{1-a_0^2}{1+a_0^2}.
\end{equation}
The branches of the spectrum (\ref{SS_spectrum}) with different chirality are implemented under the condition 
\begin{equation}\label{SS_exist}
\frac{2a_0}{1+a_0^2}\frac{mc}{\hbar} - s |\bm{k}_{||}|\geq 0.
\end{equation}

Another approach to the theory of SSs is related to the formulation of models whose bandgap parameters smoothly (at the atomic scale) vary near the interface ("heterocontact"). In this case, the multi-band method of EFs can be applied in the whole space, and no boundary condition problem arises. In [\onlinecite{Volkov_Pank_1985_English}], the authors considered interface states of DFs in the model of a symmetric inverse heterocontact. These states are described by the Dirac equation in which the mass of DFs is inverted in space, the conduction and valence bands being interchanged (Fig. 3). 

Near the inversion point, a 2D band of SSs arises. The equation for DFs at the inverse junction takes a form characteristic of supersymmetric (SUSY) quantum mechanics: 
\begin{equation}\label{Vol_Pank_eq}
\left(
\begin{array}{cc}
 -E & i\dfrac{E_g(z)}{2} + c\bm{\sigma}\cdot\bm{k} \\
-i\dfrac{E_g(z)}{2} + c\bm{\sigma}\cdot\bm{k} & -E
\end{array}
\right)
\left(
\begin{array}{cc}
\Psi_c \\
\Psi_v
\end{array}
\right)=0
\end{equation}

The zero mode of this equation describes interface states, 
\begin{equation}\label{SUSY_solution}
\Psi_{\pm} = 
\left(
\begin{array}{c}
0 \\

e^{i\theta/2}\\

\pm e^{i\theta/2}\\

0
\end{array}
\right) \exp\left\{\int_0^z \frac{E_g(z)}{2\hbar c}dz+ i\bm{k}_{||}\bm{r}_{||} \right\},
\end{equation}
where $e^{i\theta}=(k_x + i k_y)/k_{||}$. Solution (\ref{SUSY_solution}) is stable against to perturbations of the function $E_g(z)$ in view of SUSY. The Dirac point in this symmetric model is located precisely at the center of the gap, and the spectrum of SSs is actually described by formulas (\ref{SS_spectrum}) and (\ref{Dirac_point}) in which $a_0 = 1$.

%%%%%%%%%%%%%%%%%%%%%%%%%%%%%%%%%%%%%%%%%%%%%%%%%%%%%%%%%%%%%%%
\begin{figure*}[!]
\includegraphics{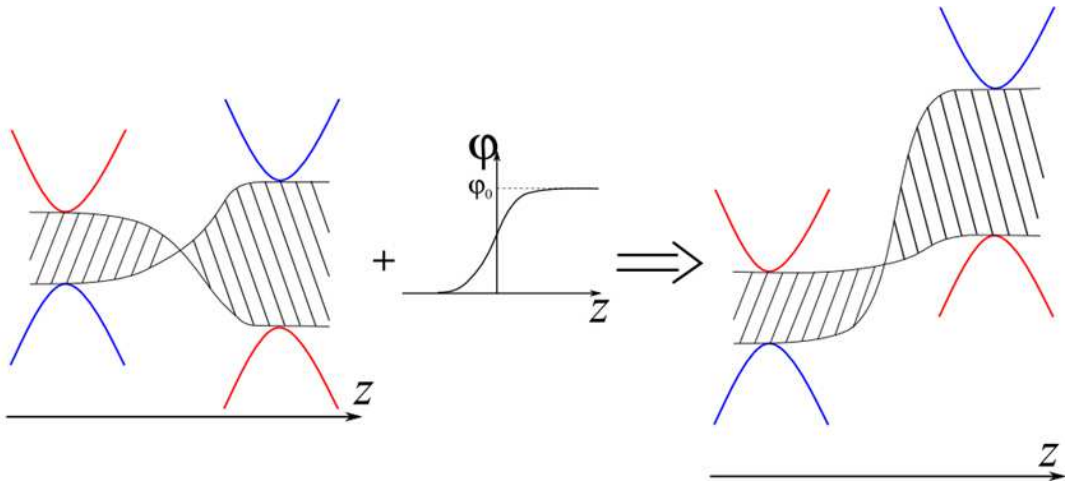}
\caption{Model of an asymmetric inverse heterocontact [\onlinecite{Volkov_review_1995_English}]. For large asymmetry, interface states cease to exist in spite of the inversion of bands. }
\end{figure*}
%%%%%%%%%%%%%%%%%%%%%%%%%%%%%%%%%%%%%%%%%%%%%%%%%%%%%%%%%%%%%%%%%

The model of an asymmetric contact considered in the survey [\onlinecite{Volkov_review_1995_English}] is more realistic (see Fig. 4). The asymmetry of bands is defined by the scalar potential $\varphi(z)$ that have the meaning of the bending of bands. 

For a greater discontinuity of bands, the spectrum of interface states is given by 
\begin{equation}\label{Volk_Pank_SS_spectra}
E=-\frac{E_{g1}\varphi_0}{E_{g0}} \pm |\bm{k}_{||}|\sqrt{1 - \left(\frac{2\varphi_0}{E_{g0}}\right)^2}
\end{equation}
where $\varphi =\varphi_0 f(z) $, $E_g(z)=E_{g0}f(z)+E_{g1}$, and the model function $f(z)$ describes a correlated variation of the band gap and the bending of bands at the heterointerface. 

It makes sense to compare the results of [\onlinecite{Volkov_Pinsker_1981_English}] and [\onlinecite{Volkov_review_1995_English}]. The heterointerface spectrum (\ref{Volk_Pank_SS_spectra}) turns into the spectrum of SSs (\ref{SS_spectrum}) under the change 
\begin{equation}
a_0 = -{\rm sign}\left(E_{g0}\right)\sqrt{\frac{E_{g0}-\varphi_0}{E_{g0}+\varphi_0}}
\end{equation}
This demonstrates the physical meaning of the boundary parameter $a_0$: its sign correlates with the sign of the gap, and the difference of its amplitude from 1 is the measure of the e--h asymmetry. An analog of the topologically non-trivial phase in Fig. 2a corresponds to the inversion of bands. An important consequence of (\ref{Volk_Pank_SS_spectra}) is as follows: in general, the inversion of bands is insufficient for the emergence of supersymmetric interface states. For higher e--h asymmetry, when $\varphi_0/E_{g0}>1$, these states cease to exist. 
%%%%%%%%%%%%%%%%%%%%%%%%%%%%%%%%%%%%%%%%%%%%%%%%%%%%%%%%%%%%%%%
\begin{figure}[!]
\includegraphics{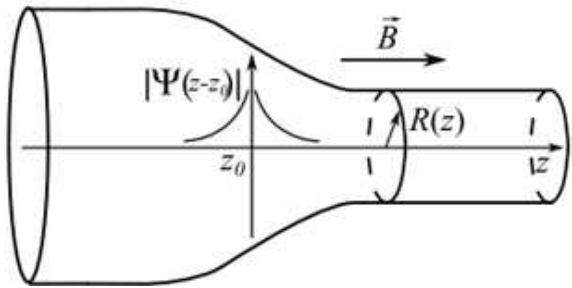}
\caption{A wire with inhomogeneous cross section in a longitudinal magnetic field. The amplitude of a supersymmetric SS that is implemented for half-integer values of a local magnetic flux (in units of flux quanta) passing through the wire cross section $z = z_0$. }
\end{figure}
%%%%%%%%%%%%%%%%%%%%%%%%%%%%%%%%%%%%%%%%%%%%%%%%%%%%%%%%%%%%%%%%%

In principle, the conditions for the emergence of SUSY states can be controlled by external fields. For example, consider [\onlinecite{Enaldiev_arxiv}] a solution to the Dirac equation (\ref{Dirac_eq}) with the boundary conditions (\ref{Dirac_BC})  for a cylindrical nanowire of radius $R_0$ in a longitudinal magnetic field $B$. The spectrum of SSs is controlled by magnetic flux $\Phi=\pi B R_0^2$ through a section of the wire. For the positive sign of the boundary parameter $a_0$ in a strong field, the quasi-classical spectrum of SSs is given by 
\begin{equation}\label{SS_wire}
E=\pm\frac{2a_0c\hbar}{1+a_0^2}\sqrt{k_z^2+\frac{(j+\Phi/\Phi_0)^2}{R^2}}+E_0,
\end{equation}
where $k_z$ is a projection of quasi-momentum to the wire axis, $j$ are half-integer numbers, and $\Phi_0 = hc/e $ is the flux quantum. When the magnetic flux is a half multiple of the flux quantum, the spectrum (\ref{SS_wire}) becomes gapless. For any sign of $a_0$, the contribution of SSs to the total density of states oscillates as a function of the magnetic flux with period $\Phi_0$. This leads to Aharonov--Bohm-type oscillations of the wire resistance. 

More interesting is the case of a wire with inhomogeneous cross section in a uniform longitudinal magnetic field. Magnetic flux through the section of a wire is inhomogeneous, whereby local SUSY states can be formed (Fig. 5).

In [\onlinecite{Enaldiev_arxiv}], it was found that such a SUSY solution with energy $E_0$ has the form 
\begin{equation}\label{SUSY_wire}
\Psi \propto \exp\left( \frac{1}{\hbar v}\int_{z_0}^{z}\Delta(z)dz\right),
\end{equation}
where $v=2a_0c/(1+a_0^2)$ is the velocity of surface states. The effective local gap $\Delta(z)=\hbar v \left(j+ \Phi(z)/\Phi_0\right)/R(z)$ is inverted at the point $z = z_0$; therefore, solution (\ref{SUSY_wire}) is localized precisely at this point.

\section{TOPOLOGICAL SURFACE STATES}

The theory of band structure of semiconductors involving topological arguments, which were first introduced in the theory of the quantum Hall effect [\onlinecite{Thouless_1982}], gave rise to a new physical concept---a topological insulator (TI) [\onlinecite{Hasan_Kane_TI_review},\onlinecite{Qi_Zhang_review}]. In topological insulators with a strong spin--orbit interaction, the specific features of the bulk band structure are such that a band of conducting surface (or edge) states of the type shown in Figs. 1c and 2a should exist on the surface of TIs for topological reasons. This is the most fundamental prediction of topological theory. Surface DFs are non-degenerate with respect to the spin quantum number (chirality) and have a conical dispersion law, which leads to the weakening of their scattering by non-magnetic impurities. It is proved that this property is a specific feature of precisely topological surface or edge states.

To implement a TI phase, one usually should impose three qualitative conditions on the parameters of the bulk band structure: symmetry with respect to time reversal (T symmetry), strong spin--orbit interaction, and inversion of bands. As a result, topological SSs arise in the bulk gap; according to topological theory, the very fact of emergence of these states does not depend on the details of the structure of the near-surface region. The possibility of implementation of 2D TI [\onlinecite{Kane_2005}] and 3D TI [\onlinecite{Fu_Kane_2007}] phases was predicted in 2005 and 2007, respectively. 

First, consider a 1D band of edge states in a 2D crystal system. The topological invariant $Z_2$ is defined by the properties of the anti-unitary matrix $w_{nm}=\langle u_n(\bm{k})|\hat{T}|u_m(-\bm{k})\rangle$, which is composed of the matrix elements of the time-reversal operator on Bloch factors of all occupied bands [\onlinecite{Hasan_Kane_TI_review}]. The T symmetry allows one to classify all 2D systems into two groups that differ by the value of the $Z_2$ invariant $\nu = \left\{0, 1\right\}$ [\onlinecite{Kane_2005}]. A crystal is said to be in the phase of a topological (trivial) insulator if the Hamiltonian characterizing it has $\nu = 1$ ($\nu = 0$). A TI phase is distinguished by the existence of an odd number of Kramers pairs of edge states at the Fermi level in the bulk gap. It is important that any T-invariant perturbations of the Hamiltonian of the crystal that do not close the gap in the spectrum cannot change the value of the $Z_2$ invariant of the system. Topological protection of SSs in a TI is understood precisely in this sense. 

%%%%%%%%%%%%%%%%%%%%%%%%%%%%%%%%%%%%%%%%%%%%%%%%%%%%%%%%%%%%%%%
\begin{figure}[ht]
\includegraphics{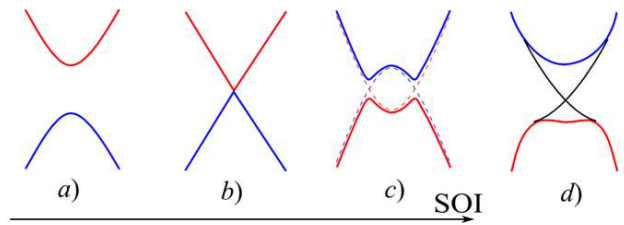}
\caption{Scheme of evolution of the bulk spectrum of a narrow-gap crystal for increasing bulk spin--orbit interaction leading to the inversion of bands: from (a) a trivial insulator phase through (b) a gapless phase to (c) the inverted spectrum in the phase of a topological insulator. (d) Bandgap model of a semi-infinite Bi$_2$Se$_3$-type topological insulator. The curvature of the spectrum of bulk bands is contributed from the dispersion the mass term $M(\bm{k})$. The states in the gap illustrate the spectrum of topological SSs. }
\end{figure}
%%%%%%%%%%%%%%%%%%%%%%%%%%%%%%%%%%%%%%%%%%%%%%%%%%%%%%%%%%%%%%%%%

In the theoretical study [\onlinecite{Bernevig_2006}], the authors predicted that a 2D TI phase should be implemented in a CdTe--HgTe double heterostructure for the width of the HgCdTe quantum well greater than 6.4 nm. It is these conditions under which the electron and hole size-quantized subbands are inverted. 

A two-component EF describing the states in a quantum well with two different values of the spin quantum number satisfies a modified Dirac equation (the so-called BHZ model [\onlinecite{Bernevig_2006}]): 
\begin{widetext}
\begin{equation}\label{BHZ}
\left(
\begin{array}{cccc}
\varepsilon(k)+m(k) & A(k_x - ik_y) & 0 & 0 \\
A(k_x + ik_y) & \varepsilon(k)-m(k) & 0 & 0 \\
0 & 0 & \varepsilon(k)+m(k) & -A(k_x + ik_y) \\
0 & 0 & -A(k_x - ik_y) & \varepsilon(k)-m(k)
\end{array}
\right)
\left(
\begin{array}{c}
\Psi_{\uparrow} \\
\Psi_{\downarrow}
\end{array}
\right)=
E\left(
\begin{array}{c}
\Psi_{\uparrow} \\
\Psi_{\downarrow}
\end{array}
\right)
\end{equation}
\end{widetext}
Here $\varepsilon(k)= C -D(k_x^2 + k_y^2)$, $m(k)= M - B(k_x^2 + k_y^2)$,
$A, B, C, D$, and $M $are the parameters of the system, and $k_x$ and $k_y$ are the components of 2D momentum. The inversion of bands corresponds to a positive value of the product of parameters $MB > 0$. 

The Tamm problem in the half-plane $x > 0$ in the model of zero boundary conditions gives rise to spin-split edge states with linear dispersion. Recently, 1D conducting edge channels associated with these states have been observed experimentally [\onlinecite{Konig_2007},\onlinecite{Nowack_2013}]. However, the spectrum of edge states has not been measured. 

In [\onlinecite{Enaldiev_2015}], the authors considered the problem of sensitivity of edge states to the form of nonzero boundary conditions. They derived a general T-invariant boundary condition for the EF $\Psi=\left(\Psi_{\uparrow},\Psi_{\downarrow}\right)^T$ satisfying Eq. (\ref{BHZ}): 
\begin{equation}\label{Zag_BC}
\left[F\partial_{\bm{n}}\Psi + G\Psi\right]_{\Gamma}=0
\end{equation}
where $G$, a matrix that mixes different components of the EF, is defined by six real phenomenological parameters and the matrix $F$ is defined by the bulk parameters $B, A$, and $D$; this matrix is introduced into the boundary condition (\ref{Zag_BC}) for convenience. The spatial symmetry reduces the number of independent phenomenological parameters in the matrix $G$ to four ones. The authors showed that the difference of boundary conditions from zero conditions significantly changes the spectrum of edge states and, generally speaking, can even lead to the vanishing of these states from the region of small quasi-momenta, where model (\ref{BHZ}) is applicable. 

Note the following important feature of this system. Although non-zero values of the parameters $B$ and $D$ are assumed to be important for topological classification, they make a small contribution to the energy of size-quantized subbands for actual quantum wells of Hg(Cd)Te [\onlinecite{Bernevig_2006}]. Therefore, one can neglect these parameters in the Tamm problem, and Eq. (\ref{BHZ}) reduces to a 2D version of the Dirac equation. For the latter equation, the spectrum of the Tamm problem is defined by formula (\ref{SS_spectrum}) in which $\bm{k}_{||}$ is an conserving projection of quasimomentum to the 1D boundary of the 2D system considered. In this case, information on the inversion of bands is contained in the sign of the boundary parameter $a_0$ rather than in the bulk parameters of the BHZ equation (\ref{BHZ}). 

The topological classification is complicated as the dimensionality increases. For a 3D TI, one usually defines four topological invariants ($\nu_0;\nu_1,\nu_2,\nu_3$), each of which can take values 0 or 1. A 3D crystal characterized by $\nu_0=1$ is called a strong TI. In the simplest case, a gapless 2D band of non-degenerate massless DFs is formed on the surface of a strong 3D TI. 3D crystals with $\nu=0$ are called weak TIs for which the appearance of SSs in the forbidden band depends on both the values of $\nu_{1,2,3}$ and the translational symmetry of the surface [\onlinecite{Fu_2007}]. In the general case, weak TIs have no gapless SSs. 

It has been theoretically shown that a whole class of  Bi$_2$Se$_3$-type compounds with inversion symmetry should belong to the TI phase. In this case, it is the strong spin--orbit interaction that leads to the inversion of two bands with different parity that are nearest to the Fermi level (Fig. 6). The existence of SSs in Bi$_2$Se$_3$, Bi$_2$Te$_3$, and Sb$_2$Te$_3$ compounds was experimentally established in [\onlinecite{Hsieh_PRL_2009},\onlinecite{Hsieh_Hasan_Nature_2009}]. A standard 3D TI for ARPES measurements is the crystalline compound Bi$_2$Se$_3$ in which the gap is about 0.3 eV. By this method, it was established [\onlinecite{Hsieh_PRL_2009}] that on the (111) surface there exist non-degenerate SSs with massless Dirac spectrum in the gap of the crystal.

In the approximation of EFs, these compounds are described near the center of the Brillouin zone by a modified Dirac equation, which, with neglect of the anisotropy and the e--h asymmetry, has the form 
\begin{equation}\label{3D_TI}
\left(
\begin{array}{cc}
M(\bm{k}) - E &  v\bm{\sigma}\cdot\bm{k} \\
v\bm{\sigma}\cdot\bm{k} & -M(\bm{k}) - E
\end{array}
\right)
\left(
\begin{array}{cc}
\Psi_c \\
\Psi_v
\end{array}
\right)=0
\end{equation}
where the significantly dispersive mass term $M(\bm{k}) = M_0 - B\bm{k}^2$ distinguishes this equation from the true Dirac equation. A topological phase is characterized by the positive value of the product of parameters $M_0B > 0$, which is attributed to the inversion of bands. The break of the e--h symmetry is described by the introduction of an additional diagonal contribution to Eq. (\ref{3D_TI}). We will neglect this asymmetry. 

To describe SSs within Eq. (\ref{3D_TI}), one should specify boundary conditions that take into account the atomic-scale structure of the surface. To this end, one often uses the fact that the above-mentioned modification of the Dirac equation doubles the order of the system of differential equations. Therefore, it is simplest to use zero boundary conditions for all components of EFs [\onlinecite{Liu_Zhang_PRB_2010}, \onlinecite{Shan_2010}]: $\left[\Psi_c = \Psi_v\right]_S = 0$. Such boundary conditions, which are often called ''open'' do not take into account the details of the microscopic structure of the surface at the atomic scale, but nevertheless guarantee the existence of massless DFs on a plane surface. 

However, it was demonstrated that non-zero boundary conditions may significantly affect the spectrum of SSs in the case of 3D TIs as well [\onlinecite{Enaldiev_2015}, \onlinecite{Menshov_2014}]. General boundary conditions for a four-component EF mix the values of the EF with its normal derivative on the boundary $S$: 
\begin{equation}
\left[\frac{B}{M_0}\partial_{\bm{n}}\Psi+Q\Psi\right]_S=0
\end{equation}
where the $4\times 4$ matrix $Q$ should be determined from additional considerations. The T symmetry, along with the spatial symmetry, reduces the number of independent real phenomenological parameters in the matrix to three ones. For sufficiently strong surface spin--orbit interaction described by these parameters, SSs may disappear from the forbidden gap in the neighbourhood of the point $\Gamma$ at which the EF method is applicable [\onlinecite{Enaldiev_2015}]. 

Experiments have shown that the properties of SSs are sensitive to the perturbations of the surface potential, which can be taken into account by an appropriate choice of the parameters of the boundary conditions. For example, the doping of the surface with non-magnetic and magnetic atoms shifts the position of the Dirac point and modifies the dispersion of SSs [\onlinecite{Valla_2012, Scholz_2012, Bianchi_2011, Roy_2014}]. 

Another Dirac system is given by a topological crystalline insulator (TCI). In a TCI, the protection of SSs is guaranteed by two symmetries at once: the T symmetry and the spatial symmetry of the crystal and its surface [\onlinecite{Fu_2011}]. A topological classification is performed similar to the above-considered case of TIs, with the difference that the matrix $w_{nm}$ is constructed on the operator $\hat{U}\hat{T}$, where $\hat{U}$ is the operator of spatial symmetry. 

Lead chalcogenides and Pb$_{1-x}$Sn$_x$Te(Se) solid solutions pass to the TCI phase at certain values of the concentration of tin when the inversion of bands of different parities occurs at the $L$ points [\onlinecite{Hsieh_NatCom_2012, Xu_2012, Dziawa_2012, Egorova_2015}]. The role of spatial symmetry in these compounds is played by mirror symmetry, which persists not for any orientations of the surface. In the Brillouin zone of this system, there are four inequivalent $L$ valleys in the forbidden band of each of which massless surface 2D DFs appear. On a (001) surface, pairs of Dirac cones of SSs are projected to a single point, leading to the degeneracy of the spectra of SSs. Atomically abrupt perturbations preserving mirror symmetry lift this degeneracy, moving apart the Dirac points in the reciprocal space without opening a gap in the spectrum of SSs [\onlinecite{Hsieh_NatCom_2012, Xu_2012, Dziawa_2012}]. However, perturbations breaking the mirror symmetry of the surface give rise to a gap in the spectrum of SSs [\onlinecite{Zeljkovic_2015}], thus breaking the topological protection. The fundamental features of the spectrum of SSs in any phase of TCIs based on lead chalcogenides can be found when one neglects the intervalley interaction in the solution of an anisotropic Dirac equation with boundary conditions (\ref{Dirac_BC}).

\section{EDGE STATES IN GRAPHENE}

The Tamm--Shockley states arise far from always. The following questions are fundamental: Under what conditions, except for those implemented in TIs, do Tamm--Shockley-type states arise? Do they exist on real interfaces under normal conditions (rather than in vacuum)? Does the conductivity through these states have a band character? The analysis carried out above shows that, in a number of Dirac materials, such states exist (in contrast to, for example, GaAs-type semiconductors). 

%%%%%%%%%%%%%%%%%%%%%%%%%%%%%%%%%%%%%%%%%%%%%%%%%%%%%%%%%%%%%%%
\begin{figure}[ht]
\includegraphics{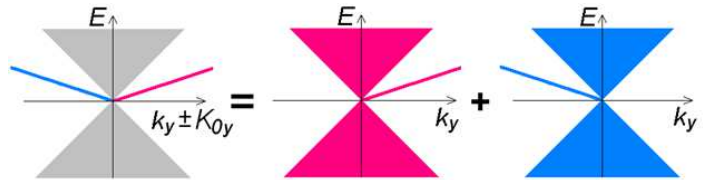}
\caption{Edge states on a graphene half-plane in each valley (on the right of the equality sign) and in the reduced valley scheme, when their centers coincide (on the left). }
\end{figure}
%%%%%%%%%%%%%%%%%%%%%%%%%%%%%%%%%%%%%%%%%%%%%%%%%%%%%%%%%%%%%%%%%

Quite recently, edge states of this type have been experimentally observed by resonance methods in another 2D Dirac material, graphene. It was found [\onlinecite{Latyshev_2013}] that the spectrum of Tamm--Shockley edge states in graphene is linear with respect to momentum, which is similar to a section of the conical spectrum of SSs in TIs. Moreover, the band character of the conductivity through these states in normal conditions [\onlinecite{Latyshev_SciRep_2014}] was proved by a direct transport experiment. Let us describe a theory necessary to explain these experiments. 

2D DFs in graphene are described by a two-valley system of equations for two-component functions $\Psi_K$ and $\Psi_{K'}$ describing EFs in two valleys, $K$ and $K'$: 
\begin{equation}\label{Graphene_eq}
\left(
\begin{array}{cc}
 v \bm{\sigma}\cdot\left(\bm{k}-\bm{K}\right) & 0 \\
0 &  v \bm{\sigma}\cdot\left(\bm{k}-\bm{K'}\right)
\end{array}
\right)
\left(
\begin{array}{cc}
\Psi_K \\
\Psi_{K'}
\end{array}
\right)=E
\left(
\begin{array}{cc}
\Psi_K \\
\Psi_{K'}
\end{array}
\right)
\end{equation}
where $\bm{\sigma}=(\sigma_x,\sigma_y)$ are the Pauli matrices in standard representation and $v\approx 10^6$ m/s. 

General boundary conditions can be determined, as usual, from two requirements: the Hermiticity of Hamiltonian (\ref{Graphene_eq}) and the invariance of the boundary conditions with respect to time reversal. They have a rather complicated form [\onlinecite{Zagorodnev2011,McCann_2004,Zag_Dev_En_PRB,Akhmerov_2008,Basko_2009,Tkachov_2009}]  
\begin{equation}\label{graphene_BC}
\left[\Psi_K \sin\beta + i e^{i\varphi}\left(e^{i\gamma\bm{\sigma}\cdot\bm{n}}+\sigma_z \cos\beta\right)\sigma_z\Psi_{K'}\right]_{\Gamma}=0
\end{equation}
Here $\beta, \varphi$, and $\gamma$ are real phenomenological parameters. 

With the neglect of intervalley scattering, the system can be described by a pair of independent 2D Weyl equations with one-parameter boundary conditions [\onlinecite{Volkov_2009},\onlinecite{Ostaay_Akhmerov_PRB}] for the single-valley function $\Psi_{K(K')}=\left(\psi_{1K(K')},\psi_{2K(K')}\right)^T$: 
\begin{eqnarray}\label{graphene_single_valley}
v \bm{\sigma}\cdot\bm{p}\Psi_{K(K')}=E\Psi_{K(K')} \\
\left[\psi_{1K(K')}+ia^{\tau}\psi_{2K(K')}\right]_{\Gamma}=0
\end{eqnarray}
where $a$ is a real phenomenological parameter that determines the properties of the edge of graphene at the atomic scale and the index $\tau=\pm 1$ numbers the valleys: $\tau=1$ ($\tau=-1$) in valley $K$ ($K'$). 

The Tamm problem (\ref{graphene_single_valley}) is easily solved. The wave function of edge states is exponentially localized near the boundary. A typical localization length determined from experiment [\onlinecite{Latyshev_SciRep_2014}] is 2 nm. The value of $a$ should be determined from comparison with experiment; however, from a comparison with model microscopic calculations [\onlinecite{Ostaay_Akhmerov_PRB},\onlinecite{Koskinen_2008}], one should expect that it is small: $|a|\ll 1$.
 
For a graphene sample in the form of a half-plane, the spectrum of these states in the absence of a magnetic field has the form [\onlinecite{Basko_2009,Tkachov_2009,Volkov_2009,Ostaay_Akhmerov_PRB}] (Fig. 7): 
\begin{equation}\label{ES_spectra}
E_{\tau}(k_{||})=\frac{2a}{1+a^2}\tau v k_{||},\quad \tau k_{||}>0
\end{equation}
Here $k_{||}$ is the one-dimensional momentum of an electron along the edge, measured from the center of the valley. It is important that the sign of $k_{||}$ is determined by the number of the valley. For clarity, different valleys in the figure are shown in different colors. The localization length of an edge state is equal to $1/k_{||}$. 

In the case of a perpendicular magnetic field, two types of edge states exist: Tamm--Shockley states and magnetic edge states due to skipping orbits. They interact with each other, and the result of their interference depends on the geometry of a sample. For the half-plane $x > 0$, the resulting spectrum [\onlinecite{Zagorodnev2011}] is shown in Fig. 8. 
%%%%%%%%%%%%%%%%%%%%%%%%%%%%%%%%%%%%%%%%%%%%%%%%%%%%%%%%%%%%%%%
\begin{figure}[ht]
\includegraphics{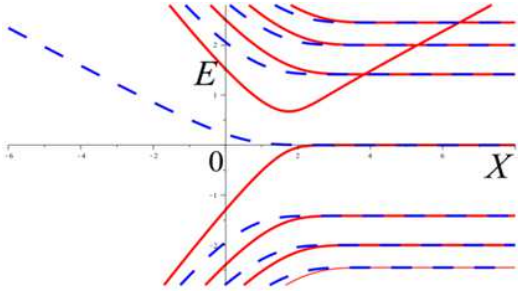}
\caption{Energy (in units of $\hbar v/\lambda$) of DFs as a function of the $X$ center of oscillator in a semi-infinite graphene in a magnetic field of 1 T for $a = 0.2$. The reduced valley scheme. Red solid lines show the states from valley $K$, and blue dashed lines, the states from valley $K'$.  }
\end{figure}
%%%%%%%%%%%%%%%%%%%%%%%%%%%%%%%%%%%%%%%%%%%%%%%%%%%%%%%%%%%%%%%%%
Edge states also exist on the boundary of a circular hole (an antidot with radius $R_0$) in an infinite sheet of graphene. The finiteness of the perimeter of the antidot leads to the quantization of the momentum component parallel to the boundary. Now discrete edge states are characterized by the total angular momentum $j = \pm 1/2, \pm 3/2, \pm 5/2, \dots$, different signs of $j$ corresponding to different valleys. Within a quasiclassical approach, the spectrum of such edge states is obtained from (\ref{ES_spectra}) by replacing $k_{||}$ by a multiple of $1/R_0$. A more accurate asymptotes ($|a| \ll 1$) yields a decay-type complex spectrum: 
\begin{equation}\label{ES_AD}
E=\frac{2a\hbar v}{R_0}\left(|j|-\frac{\tau}{2}\right) - i \frac{2\pi\hbar v}{R_0}\frac{|a|\left(|a|(|j|-1/2)\right)^{2|j|-1}}{\Gamma^2(|j|-1/2)}.
\end{equation}
Here $\tau j > 0$, $\Gamma(z)$ is a gamma function, and $j = \pm 3/2, \pm 5/2, \dots$. In a special case of $j = \pm 1/2$, the energy of edge states is determined by the root of the equation 
\begin{equation}\label{ES_zero_mode}
ER_0\ln\left(\frac{|E|R_0}{2}\right) = - a
\end{equation}

Although these states are quasistationary in the absence of a magnetic field, their lifetime with respect to the decay into bulk states is large in the actual case of small $a$. DFs trapped in edge states perform a cyclic motion along the perimeter of the antidot (clockwise in one valley, or counterclockwise in the other), acquiring an additional Aharonov--Bohm phase in a magnetic field.

The spectrum of edge states in the antidot in a perpendicular magnetic field is controlled by a magnetic flux $\Phi=\pi R_0^2B$ through the area of the antidot: 
\begin{equation}\label{ES_asymp}
E = \frac{2a\hbar v}{R_0}\left(j+\frac{\Phi}{\Phi_0}-\frac{\tau}{2}\right).
\end{equation} 
The applicability conditions of this asymptotes is given by 
\begin{equation}
\tau\left(j+\frac{\Phi}{\Phi_0}\right)>0, \quad |a|\ll 1.
\end{equation} 

Asymptotes (\ref{ES_asymp}) is the more justified, the less the effect of the magnetic field on the orbital part of the wave function, i.e., when the localization length of an edge state is small compared with the magnetic length. In the general case, a numerically calculated spectrum is shown in Fig. 9 [\onlinecite{Enaldiev2011}]. 
%%%%%%%%%%%%%%%%%%%%%%%%%%%%%%%%%%%%%%%%%%%%%%%%%%%%%%%%%%%%%%%
\begin{figure}[ht]
\includegraphics{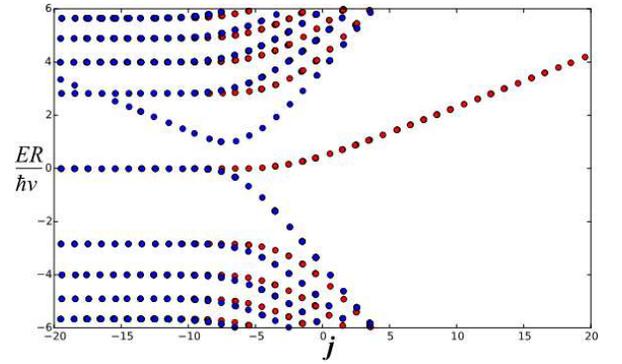}
\caption{Dimensionless energy of DFs at a graphene antidot in a magnetic field of $B = 4$ T as a function of the conserving projection of the total angular momentum $j$ in the reduced valley scheme; $a = 0.1$. The states in different valleys are shown in different colors.}
\end{figure}
%%%%%%%%%%%%%%%%%%%%%%%%%%%%%%%%%%%%%%%%%%%%%%%%%%%%%%%%%%%%%%%%%
These states explain the recently revealed Aharonov--Bohm effect in the resistivity of nanoperforated graphene. From comparison with experiment, we can determine the value of the boundary parameter $a = -0.05$. 

Thus, we can conclude that hole-type DFs slowly rotate around every antidot in nanoperforated graphene. The sign of rotation is determined by the number of the valley, and the speed of rotation is ten times less than the volume velocity of DFs in graphene. 

\section{CONCLUSIONS. MINIMAL MODEL OF SURFACE STATES FOR DIRAC ELECTRONS} 

One can draw the following qualitative conclusion from the survey of literature data on Dirac materials: the Dirac property favours the emergence of a Tamm--Shockley-type band of edge and surface states. However, this band is not always situated in the bulk gap where its spectrum is analogous to the spectrum of topological states. For some values of boundary parameters, this band is expelled from the gap and its energy overlaps with the energy of bulk bands. 

The analysis carried out allows us to formulate a minimal model suitable for the analytic investigation of fundamental features of the spectrum of surface or edge states in a number of Dirac systems: a generally anisotropic version of the 3D Dirac equation (\ref{Dirac_eq}) with boundary conditions (\ref{Dirac_BC}) and their 2D generalizations. 

Such a model can be applied with neglect of intervalley interaction to describe SSs in the following Dirac systems. 

In 3D systems: lead chalcogenides and solid solutions based on lead chalcogenides Pb$_{1-x}$Sn$_x$Se(Te)-type, bismuth- and Bi$_{1-x}$Sb$_x$-type semimetals at the $L$ point, both in the inverted and ordinary, non-inverted, modes. 

In 2D systems: graphene and 2D TIs in quantum wells based on a HgTe--CdTe heteropair. 

The minimal model cannot be applied to describe SSs in Bi$_2$Se$_3$, Bi$_2$Te$_3$, and Sb$_2$Te$_3$ because of a significant contribution of remote bands that forms the dispersion of the mass term in a modified Dirac equation. 

\section{ACKNOWLEDGMENTS} 
We are grateful to I.V. Zagorodnev for critical remarks. 

This work was supported in part by the Russian Foundation for Basic Research.

\bibliography{thesis}

\end{document}